# Super-Hydrophobic Stearic Acid Layer Formed on Anodized High Purified Magnesium for Improving Corrosion Resistance of Biodegradable Implants

Sohrab Khalifeh[a][1], and T. David Burleigh[2]

[a] Department of Materials and Metallurgical Engineering Department, New Mexico Institute of Mining and Technology, Socorro, New Mexico, USA

**Abstract**

Magnesium and its alloys are ideal candidates for biodegradable implants. However, they can dissolve too rapidly in the human body for most applications. In this research, high purified magnesium (HP-Mg) was coated with stearic acid in order to slow the corrosion rate of magnesium in simulated body fluid at 37℃. HP-Mg was anodized to form an oxide/hydroxide layer, then it was immersed in a stearic acid solution. Electrochemical impedance spectroscopy and potentiodynamic polarization were used to estimate the corrosion rate of HP-Mg specimens. The results confirm that the hydrophobic coating can temporarily decrease the corrosion rate of HP-Mg by 1000x.

## 1. Introduction

Biodegradable magnesium is being considered for the field of biomaterials, especially for orthopedic implants and heart vascular stents [1, 2]. Magnesium is an ideal biomaterial because of biocompatibility and low stress shielding [3, 4, 5, 6, and 7]. However, magnesium and its alloys have a rapid degradation rate in the biological environment that can cause serious problems such as hydrogen bubbles, caustic burning, loss of mechanical properties, and disappearance of the implant before complete healing [8, 9, 10, 11, 12, 13]. Coating is one of the methods which is used to control the corrosion rate of magnesium [14, 15, 16, 17, 18, and 19]. As mentioned, biodegradation is one distinct advantage magnesium has as a biomaterial, therefore, the coating layer should also dissolve and be biocompatible. The corrosion rate should be reduced by coating the implant, but not be permanently stopped. Stearic acid has been selected for use as hydrophobic layer in this research. Stearic acid is a carboxylic acid, and has the chemical formula $CH_3(CH_2)_{16}COOH$ [20]. Stearic acids are non-toxic and biocompatible [21]. Also, stearic acids can cause a higher reduction in LDL cholesterol compared to other saturated fatty acids due to a lower probability of transforming stearic acids into cholesterol esters [22]. Ng et al. [23] showed that magnesium coated with stearic acid can improve the corrosion resistance of pure Mg in Hank's solution.

---

[1] Corresponding author. *Email address:* skhalife@nmt.edu
[2] burleigh@nmt.edu

Hydrophobic coatings are usually formed using at least two steps, the first step is surface preparation to ensure adequate adhesion to the hydrophobic layer, and the second step is formation the hydrophobic layer on substrate. Wang et al. [20] conducted chemical etching in $H_2SO_4$, $H_2O_2$ method as the first step. In this paper, the effect of hydrophobe stearic acid layer on the corrosion resistance of high purified magnesium (HP-Mg) *in vitro* were investigated. Anodizing and immersion were used as the first and second steps of fabrication of a hydrophobe stearic acid layer on magnesium.

## 2. Material and Methods

### 2.1. Sample Preparation

High purified magnesium (HP-Mg) was used in this research and its chemical composition is given in Table 1. The HP-Mg rods were cut into 10 mm long cylinders and cold mounted in epoxy with 2.55cm$^2$ surface area. Next, the specimen surfaces were polished with sand paper (600-1200 grit), and then with 1 $\mu$m oil-based diamond slurry, degreased with ethanol, washed with deionized water, and dried using compressed air.

*Table 1. Composition for HP-Mg used in this work. All compositions are given in weight percent.*

| Material | Mg | Al | Zn | Ca | Si | Mn | Fe | Ni | Cu | Zr | Na |
|---|---|---|---|---|---|---|---|---|---|---|---|
| HP Mg | 99.97 | 0.002 | 0.005 | 0.001 | 0.014 | 0.001 | 0.003 | 0.002 | >0.0002 | >0.002 | 0.002 |

### 2.2. Anodization Process

The anodizing processes were performed at 30, 50, and 70°C for 10 seconds, at 120 volts (AC) in the aqueous electrolyte of borate benzoate. The effect of surface preparation was investigated in this research. Two groups of HP-Mg were prepared, the first samples were ground to 1200 grit and the next group was polished to 1 $\mu$m. HP-Mg was connected to a positive terminal of the power supply and the platinum counter electrode was connected to the negative terminal. After the anodizing process, the samples were rinsed by deionized water and dried by low pressure compressed air before coating with stearic acid. The magnesium surface was white after anodization.

### 2.3. Electrolyte

The borate benzoate electrolyte was composed of 60 g/L NaOH, 25 g/L $Na_2B_4O_7$, 20 g/L $H_3BO_3$, and 3 g/L NaBz (sodium benzoate: $C_6H_5COONa$) [24]. During the anodizing experiments, the temperature of the electrolyte was held at 30°C using a hot plate.

### 2.4. Hydrophobic Layer Formation

The formation process was performed using immersion. For this purpose anodized magnesium was immersed in the 0.05, 0.1, and 0.15 mol/L stearic acid in ethanol at room temperature for an hour. After immersion, the specimens were brought out from the solution, and dried in the fume hood for 24 hours.

### 2.5. Evaluation of Hydrophobic Layer

Surface morphologies of the anodized specimen was carried out using scanning electron microscopy (SEM). The water contact angles were measured at room temperature using the Kruss Drop Shape Analyzer (DSA25) in order to evaluate the hydrophobicity of coated layer.

### 2.6. Simulated Body Fluid (SBF) Preparation

All corrosion tests were performed as *in vitro* based on the human biological environment. The samples were immersed in simulated body fluid (SBF) at 37°C. The SBF was made according to the protocol for preparing SBF as described by Kokubo [25]. The ion concentrations of SBF versus human blood plasma are shown in Table 2.

*Table 2. SBF versus human blood plasma [25].*

|  | Simulated Body Fluid (SBF)(mM) | Blood Plasma (mM) |
|---|---|---|
| $Na^+$ | 142.0 | 142.0 |
| $K^+$ | 5.0 | 5.0 |
| $Mg^+$ | 1.5 | 1.5 |
| $Ca^+$ | 2.5 | 2.5 |
| $Cl^-$ | 148.8 | 103.2 |
| $HCO_3^-$ | 4.2 | 27.0 |
| $HPO_4^{2-}$ | 1.0 | 1.0 |
| $SO_4^{-2}$ | 0.5 | 0.5 |

## 2.7. Electrochemical Measurements

Electrochemical impedance spectroscopy (EIS), and potentiodynamic polarization (PDP) of the specimens were measured in SBF at 37°C using a PARSTAT 2263 potentiostat. For all measurements, a three-electrode electrochemical cell was used, with a KCl saturated calomel electrode (SCE) as a reference electrode and platinum wire as a counter electrode. At least three tests were performed for each condition with three different specimens to confirm the reproducibility of the EIS and PDP measurements. EIS test began 5-10 minutes after the specimens were immersed in SBF at 37°C, and run in the frequency range from 100 kHz to 100 mHz.

The PDP test was run immediately after the final EIS test (about 15 minutes immersion). The initial potential was -250 mV relative to open circuit potential and stopped at +1600 mV versus SCE scan at a rate of 10 mV/s. Extrapolation of the cathodic and anodic region of Tafel behavior gave the corrosion rate, $i_{corr}$ $(A/cm^2)$ at $E_{corr}$.

Polarization resistance, $R_p$ was determined from the EIS Bode plot. The impedance at 100 mHz ($R_p + R_s$) was reduced by the solution resistance at 100 kHz ($R_s$) to obtain $R_p$ (polarization resistance). Then, $R_p$ was converted to the corrosion rate, $i_{corr}$ $(A/cm^2)$ by the following equation [26];

$$i_{corr} = \frac{B}{R_p} = \frac{\beta_a \beta_c}{2.3 (\beta_a + \beta_c) R_p} \qquad (1)$$

Where $\beta_a$ and $\beta_c$ are anodic and cathodic Tafel constants, respectively, and $B$ is proportionality constant, which was determined from PDP electrochemical data. $i_{corr}$ $(A/cm^2)$ from each EIS and PDP test was converted to the corrosion rate mm/year as a $C.R._{EIS/PDP}$ by the following calculation;

$$C.R._{EIS/PDP} = i_{corr} \left(\frac{A}{cm^2}\right)\left(\frac{C}{A \cdot sec}\right)\left(\frac{e^{-1}}{1.602 \times 10^{-19} C}\right)\left(\frac{Mg\ atom}{2e^-}\right)\left(\frac{mol\ Mg}{6.022 \times 10^{23}\ Mg\ atom}\right)$$

$$\left(\frac{24.305g\ Mg}{mol\ Mg}\right)\left(\frac{cm^3}{1.74g}\right)\left(\frac{10mm}{cm}\right)\left(\frac{3600sec}{h}\right)\left(\frac{24h}{day}\right)\left(\frac{365day}{year}\right) \qquad (2)$$

$$= 22.83 \times 10^3\ i_{corr}\ mm \cdot year^{-1}$$

## 3. Results

### 3.1. Microstructure

The anodization processes were conducted using 4, 5, and 6 volts (DC) and 120 volts (AC), but only the 120 volts (AC) data is shown here. With the naked eye, the anodized surface was white after the AC current. Specimens were ground to 1200 grit, anodized at 30°C for 10 seconds in borate benzoate. Figure 1 illustrates the SEM images of the anodized specimens at different magnifications. Figure 1-c and d show that the anodization process provided the tubular morphology on magnesium substrate.

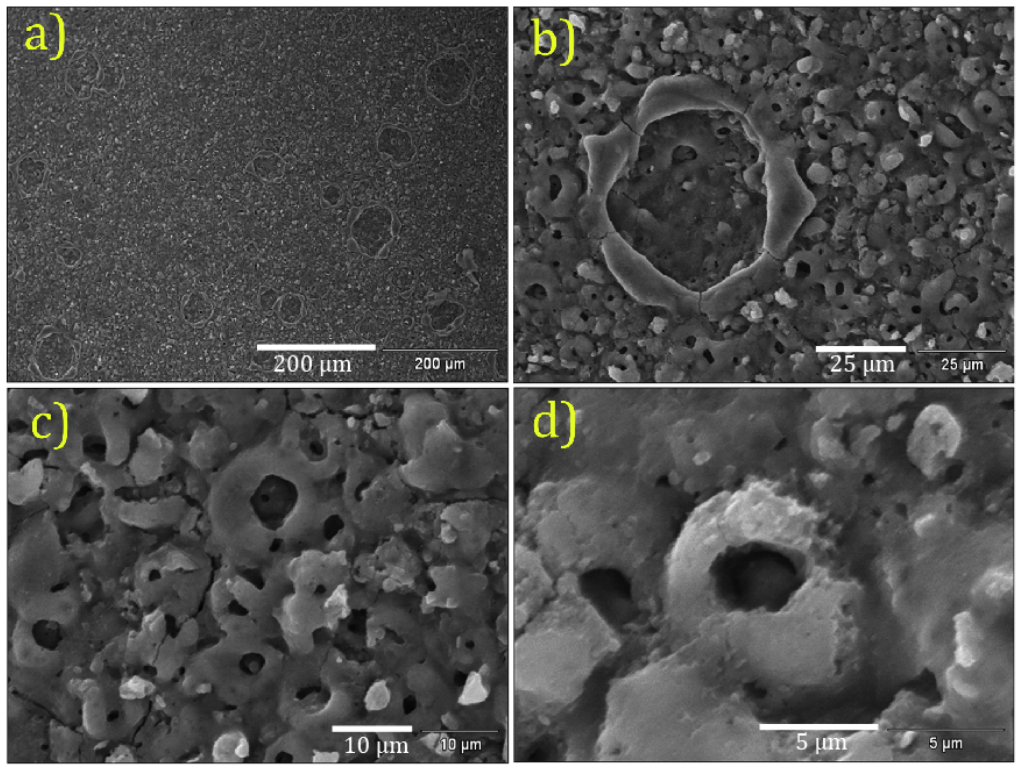

*Figure 1. SEM images of anodized HP-Mg at different magnifications.*

### 3.2. Hydrophobicity (Contact Angle)

The water contact angles were measured in order to determine the hydrophobicity of the coated specimens. Figure 2 illustrates the drop shape of the water on the polished bare HP-Mg substrate and coated HP-Mg with stearic acid. This coated HP-Mg was polished to 1200 grit, anodized for 120 volts for 10 seconds at 30°C in borate benzoate, and immersed in 0.05 mol stearic acid in ethanol for 60 minutes. Figure 2-b shows that the coating layer provided the super hydrophobicity (165°) in HP-Mg compared to the polished bare HP-Mg.

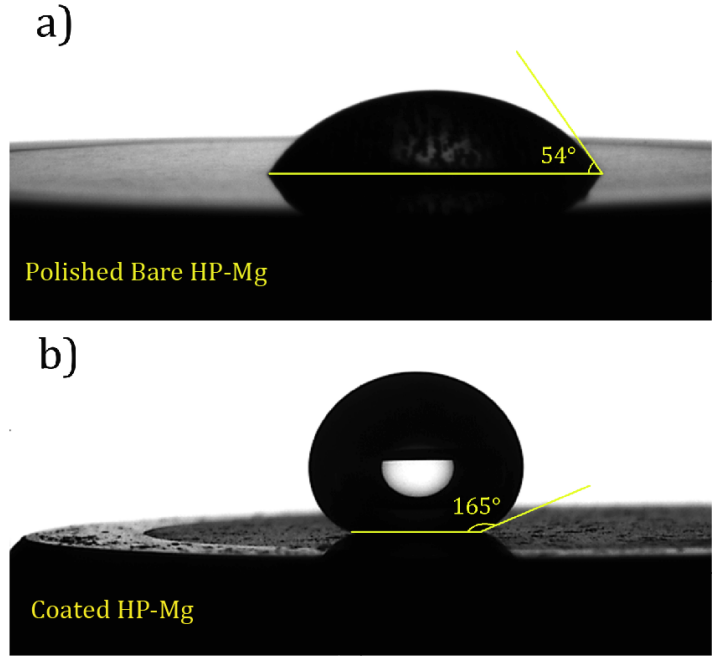

*Figure 2. The water contact angle of the drop shape of the water, a) Polished bare HP-Mg to 1200 grit , b) Coated HP-Mg, polished to 1200 grit, anodized for 120 volts and 10 seconds at 30°C in borate benzoate, and immersed in 0.05 mol stearic acid in ethanol for 60 minutes.*

### 3.3. PDP and EIS Results

In order to estimate the corrosion rate of coated HP-Mg, EIS and PDP methods were performed in SBF at 37°C. Figure 3 and 4 illustrate the corrosion rate (mm/year) based on the EIS test and PDP results of polished bare HP-Mg and coated HP-Mg, respectively. The conversion constant, B, was estimated to be 0.07 V based on our PDP and EIS experiments. The coated specimens were anodized at different temperatures (30, 50, and 70°C). The EIS and PDP results confirm that the corrosion rate of HP-Mg was significantly dropped by fabrication of the hydrophobic coating on the magnesium. Increasing the temperature of the borate benzoate electrolyte, also improved the corrosion resistance of HP-Mg.

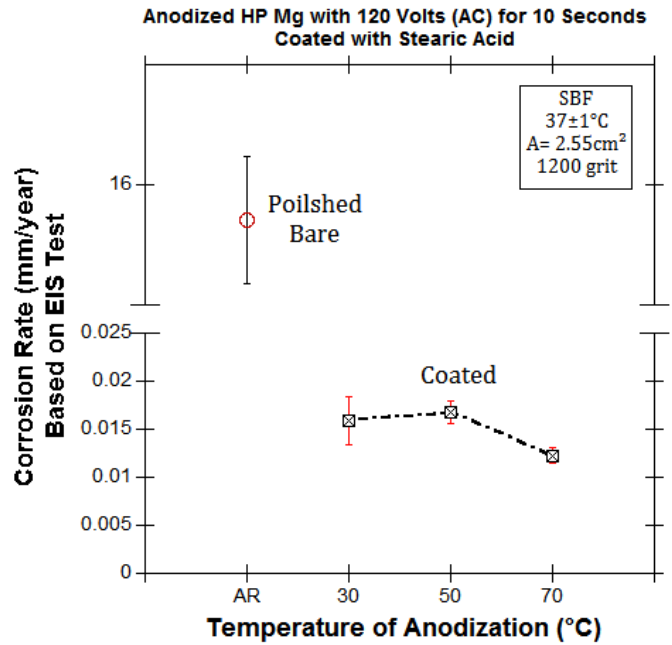

Figure 3. The effect of the anodizing temperature on corrosion rate (mm/year) of HP-Mg based on EIS test in SBF at 37°C compared to polished bare HP-Mg.

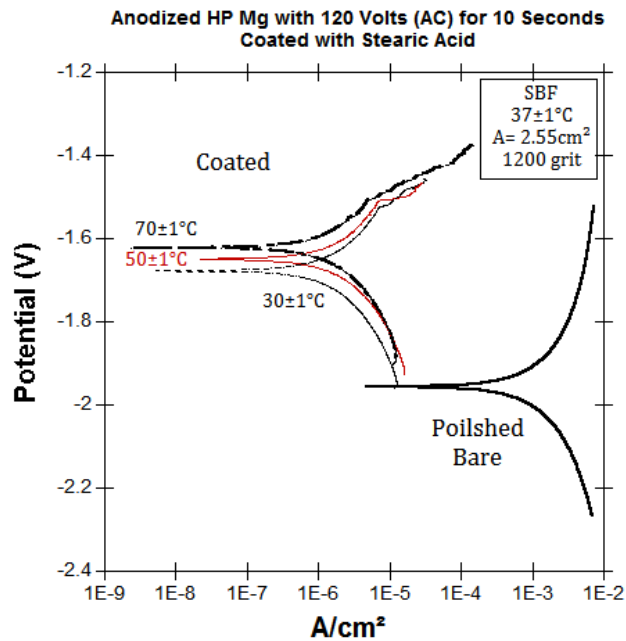

Figure 4. The effect of anodizing temperature of coating process on PDP results of coated and polished bare HP-Mg in SBF at 37°C.

Figure 5 shows the effect of changing the concentration of stearic acid solution on the corrosion rate (mm/year) of coated HP-Mg based on the EIS and PDP tests. Figure 5 illustrates that the change in the corrosion rate was negligible.

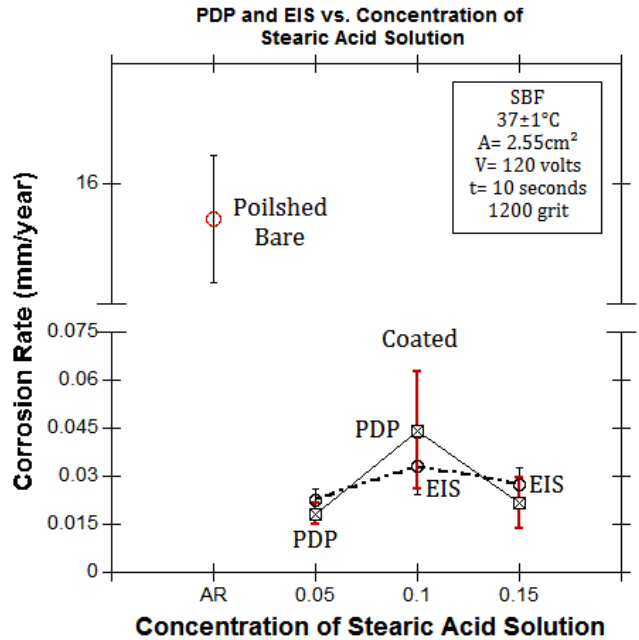

*Figure 5. The effect of stearic acid concentration on the corrosion rate (mm/year) of HP-Mg based on EIS and PDP tests in SBF at 37°C.*

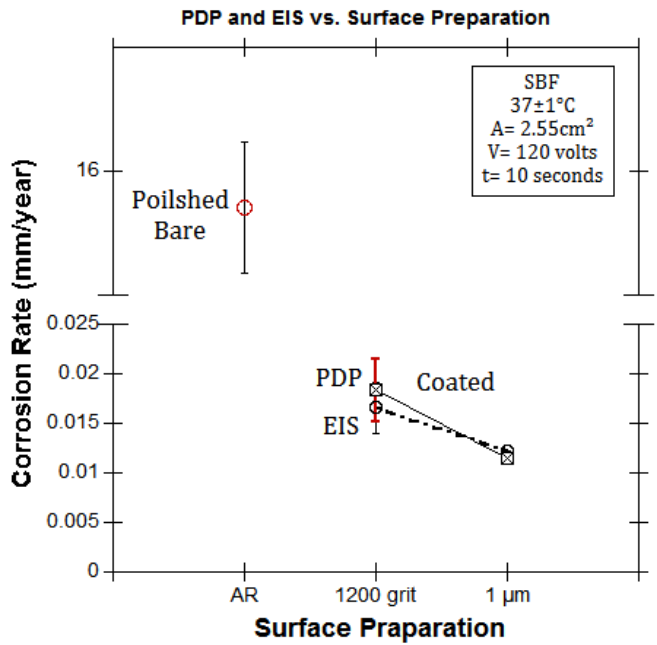

*Figure 6. The effect of surface preparation on anodization process on the corrosion rate (mm/year) of HP-Mg based on EIS and PDP tests in SBF at 37°C.*

The effect of surface preparation on the corrosion rate of HP-Mg is shown in Figure 6 based on the EIS and PDP tests in SBF at 37℃. The 1 $\mu$m diamond polish provided the better base layer for anodization.

## 4. Discussion

As mentioned before, biodegradation and biocompatibility are some advantages of magnesium when used as an orthopedic implant, along with less stress shielding and good strength to weight ratio. However, the poor corrosion resistance of magnesium in a biological environment confines the use of magnesium as a biodegradable implant. Slowing the corrosion rate of magnesium by coating with biocompatible stearic acid has been investigated in this research. We demonstrated that fabrication of the hydrophobic layer on HP-Mg can drop the corrosion rate (mm/year) by 1000x. Two electrochemical techniques were conducted to measure the corrosion rate of HP-Mg. The EIS and PDP results confirm each other and illustrated that the measured corrosion rates in this research were consistent.

We demonstrated that the coated layer obtained super hydrophobicity (Figure 2-b) in HP-Mg. Based on the SEM images (Figure 1), the tubular structure formed by anodization process could be an appropriate substrate for the stearic acid to form a hydrophobic layer.

The results from different conditions of anodization (electrolyte temperature and surface preparation) and immersion in stearic acid (different concentration of stearic acid in ethanol) (Figure 3, 6, and 5, respectively) illustrated that all anodizing processes with 120 volts (AC) for 10 seconds can reduce the corrosion rate of magnesium 1000x. Increasing the temperature of anodization to 70℃ obtained twice the improvement compared to 30℃. This might be due to the higher porosity of the magnesium oxide/hydroxide layer, and consequently more surface of the magnesium substrate to bind with stearic acid [27].

Figure 5 shows that different concentrations of stearic acid did not affect the corrosion rate of magnesium. The specimen immersed in 0.1 mol/L stearic acid in ethanol showed a higher corrosion rate compared to 0.05 mol/L solution, but increasing the concentration to 0.15 mol/L, the corrosion rate reduced and it was close to the results from 0.05 mol/L. It seems that the amount of 0.05 mol/L stearic acid in ethanol was enough to form the hydrophobic layer on magnesium substrate and the excess amount precipitates on hydrophobic layer, which would be dissolved in SBF during the EIS and PDP tests. Therefore, the excess concentration of stearic acid was not more effective on improving the corrosion behavior of HP-Mg.

The influence of surface preparation was similar as the effect of anodizing temperature of coating process on the corrosion rate of HP-Mg. The 1 $\mu$m polish enhanced the corrosion resistance twice over the 1200 grit polished specimens. The smooth surface has higher efficiency in anodizing process to provide the porous oxide/hydroxide magnesium layer.

The EIS results at the long term test showed that the coating layer broke down after 12-24 hours immersion in SBF at 37℃, and the corrosion resistance was reduced. Future work will determine ways to extend the coating protectiveness.

## 5. Conclusions

A hydrophobic layer was formed on HP-Mg in order to increase the corrosion resistance of magnesium for use as a biodegradable magnesium implant. Formation of a hydrophobic layer by anodization at 120 volts for 10 seconds was effective for improving the corrosion resistance of HP-Mg. The results affirm the following:

The formation of the stearic acid layer on HP-Mg substrate provided super hydrophobicity (165°). The corrosion rate of HP-Mg was dropped from 15 mm/year down to 0.018 mm/year by the hydrophobic layer formation. The optimum process was anodizing at 120 volts (AC) for 10 seconds, then immersion in 0.05 mol/L stearic acid and in ethanol for 60 minutes. The corrosion rates of HP-Mg were progressively reduced from 0.018 mm/year down to 0.011 mm/year by increasing the anodizing temperature to 70℃. The change in concentration of stearic acid did not affect the corrosion behavior of HP-Mg. The 1 $\mu$m polished specimens showed the lower corrosion rate (0.01 mm/year) versus the 1200 grit polished specimens. Long term results showed that the coating layer was a temporary layer and its effectiveness disappeared after 12-24 hours.